\documentclass[%
 aip,
 amsmath,amssymb,
 reprint,%
]{revtex4-2}

\usepackage{graphicx}
\usepackage{dcolumn}
\usepackage{bm}

\usepackage[utf8]{inputenc}
\usepackage[T1]{fontenc}
\usepackage{mathptmx}
\usepackage{etoolbox}

\usepackage{braket}
\usepackage[hidelinks]{hyperref}

\makeatletter
\def\@email#1#2{%
 \endgroup
 \patchcmd{\titleblock@produce}
  {\frontmatter@RRAPformat}
  {\frontmatter@RRAPformat{\produce@RRAP{*#1\href{mailto:#2}{#2}}}\frontmatter@RRAPformat}
  {}{}
}%
\makeatother

\begin{document}


\title{Adaptive Sampling in Low-Signal 2D Fourier-Transform Spectroscopy}%

\author{Sebastian Stadler} 
\author{Simon Kehrer} 
\author{Thorsten Schumacher} 

\author{Markus Lippitz} 
\email{markus.lippitz@uni-bayreuth.de}
\affiliation{Experimental Physics III, University of Bayreuth, 95447 Bayreuth, Germany}

\date{\today} 

\begin{abstract}
In Fourier transform spectroscopy, time-domain interferograms are recorded to obtain the desired spectrum by Fourier transformation. The quality of the spectrum depends on the number and accuracy of the time-domain data points. This is particularly relevant for weak and noisy signals when the time available for an experiment is limited, for example due to photobleaching or sample stability in general. In such cases, experimenters must decide whether to invest integration time in determining more Fourier coefficients or fewer coefficients more accurately. 

We present an adaptive algorithm that determines which coefficient to measure next based on previously acquired data and a simple surrogate model. This algorithm performs close to optimally across a wide range of experimental conditions, and can be easily implemented. To demonstrate its performance, we use the example of fluorescence-detected 2D spectroscopy of a single molecule. The commonly used sampling scheme of measuring a fixed set of coefficients with fixed integration times only performs optimally within a narrow range of experimental conditions. In our simulations, the adaptive algorithm reduces the mean squared error by approximately a factor of 30 compared to our previous approach.

\end{abstract}

\maketitle 

\section{Introduction}
  
Two-dimensional electronic spectroscopy (2DES) is a nonlinear spectroscopic technique which uses a sequence of ultrashort laser pulses to study coherent and incoherent ultrafast processes on the femtosecond timescale \cite{Fresch2023}. While coherent radiation from the sample has traditionally been detected, the action of the excitation process — such as fluorescence, photocurrent or photoionisation — has recently emerged as another suitable observable \cite{Bolzonello2026}. This enabled us to achieve the ultimate sensitivity limit and study a single fluorescent molecule \cite{Jana2024}.  

A single dye molecule at room temperature allows the detection of about $10^6$ photons before it inevitably photobleaches. One has to carefully consider how to sample the three time dimensions $s(t_1, t_2, t_3)$ between the four laser pulses from which the 2D spectrum $S(\omega_1, t_2, \omega_3)$ is obtained by Fourier transformation. To this end, concepts of sparse sampling have been applied to 2D spectroscopy \cite{Roeding2017,Wang2020,Bolzonello2024}. For an introduction to the topic and a recent review see  Bolzonello et al. \cite{Bolzonello2026}.

For sparse sampling to be applicable to single-molecule 2D spectroscopy, another crucial aspect must be considered: the signal-to-noise ratio. To our knowledge, all previous studies on sparse sampling in 2D spectroscopy have only discussed the low noise limit (Wang et al. \cite{Wang2020} assume a signal-to-noise ratio of 100). 
At low noise levels, increasing the integration time per time-domain point $(t_1, t_2, t_3)$ does not typically reduce uncertainty, since successive measurements are usually correlated due to drift or imperfect calibration. However, in an experiment involving only $10^6$ detection events, noise is primarily driven by uncorrelated shot noise, which can be mitigated through repeated measurements.
Therefore, in addition to deciding which points to measure, as in sparse sampling, we also need to decide how long to integrate at each point to reduce its noise. Would it be better to measure more points with a shorter integration time (i.e. more noise per point) or fewer points with a longer integration time (i.e. less noise per point)?

Furthermore, it is impossible to predict when the molecule will bleach. The measurement scheme cannot be based on a fixed total integration time. An ideal scheme would produce optimal results at any instant of photobleaching. Lastly, as the strength of the nonlinear signal and the signal-to-noise ratio obtained vary between individual molecules, the measurement scheme must be adaptable.

Below, we present an adaptive sampling algorithm that acquires 2D spectra in an almost optimal manner. Our aim is to determine the full, purely absorptive 2D spectrum. In this unexplored field of single-particle 2D spectroscopy, we avoid making strong assumptions about the measured spectrum by recovering only derived parameters such as peak positions or amplitudes. We will  use numerical simulations to compare the algorithm to both the ideal case and typical non-adaptive experimental schemes.

\section{Adaptive sampling}

We assume experimental conditions as in Jana et al. \cite{Jana2024,Jana2025}: We employ a rapid phase-cycling scheme, measure in a rotating frame and use lock-in detection to obtain complex-valued Fourier coefficients. The noise is dominated by photon shot noise, and is constant across all measured time-domain points $(t_1, t_2, t_3)$, irrespective of the value of the measurement \cite{Jana2024}.

Our metric for the quality of a measurement scheme is the mean squared error $\chi^2$, between complex-valued true ($S$) and measured ($S'$) spectrum, in a given spectral range $\pm \Omega$ around the rotating frame frequency (for simplicity written here in one dimension)
 \begin{equation}
 \chi^2 = \frac{1}{2\Omega}  \int_{-\Omega}^{\Omega} \,  \left| S(\omega)  -  S'(\omega) \right|^2 d\omega
  = \sum_{n=-\infty}^\infty \left| c_n - c'_n \right|^2 \quad .
  \label{chisq}
\end{equation}
The second equality follows from Parseval's theorem. The $c_n$ ($c'_n$) are the true (measured) complex-valued Fourier coefficients of the spectrum. The  spectral range $\Omega$ defines the distance $T = 1 / 2\Omega$ of the points in time domain.

Let us assume for a moment a low noise level. If a coefficient $n$ is measured, it contributes to $\chi^2$ with its mean squared error $ \chi^2_n = \left\langle |c_n - c'_n|^2 \right\rangle = \sigma^2$, if it is not measured, it contributes with its true value $\chi^2_n = |c_n|^2$. Since we aim to minimize $\chi^2$, the cost function becomes
\begin{equation}
   \sum_{n} \chi^2_n = \sum_{n, y_n = 0}  |c_n|^2 + \sum_{n, y_n = 1}  \sigma^2 \quad ,
   \label{eq:costHighSNR}
\end{equation}
where $y_n$ is a binary vector that defines whether a coefficient is measured ($y_n=1$) or not ($y_n=0$). This is equivalent to the sparse sampling LASSO method \cite{Brunton_buch,Bolzonello2026}, where one aims at minimizing $\chi^2$ by measuring only a subset of coefficients. The noise $\sigma^2$ is the regularisation parameter $\lambda$ of the LASSO method.
Minimizing eq.~\eqref{eq:costHighSNR} could be done by sorting all coefficients by their absolute value, and measuring the largest ones until the absolute value of the coefficients drops below the noise $\sigma^2$.  For a practical scheme, this requires a good estimate of the absolute value of the true coefficients.

In the case of high noise levels, we can increase the accuracy of a measurement by increasing the integration time or measuring it multiple times. The binary vector $y_n$ is replaced by a vector of $N_n$ that defines how many times a coefficient is measured (at the moment we do not require $N_n$ to be integer). The mean squared error of a measured coefficient is then $\chi^2_n = \sigma^2/N_n$. The cost function becomes
\begin{equation} \label{eq:MSE_target}
     \left. \left( \sum_{n, N_n = 0} |c_n|^2 + 
     \sum_{n, N_n > 0}
      \frac{\sigma^2}{N_n}  \right)\right|_{N_\text{tot} = \sum_n N_n = \text{const.}}
   \end{equation}
with a constraint on the total integration time $N_\text{tot}$. By converting the constraint to a Lagrange multiplier, one finds that the optimal solution contains $N_n = N_\text{tot}/m_\text{opt}$ where $m_\text{opt}$ is the number of unique measured coefficients, i.e., all measured coefficients are measured the same number of times.

The optimal choice of $m_\text{opt}$ is such that the addition of one more coefficient to the measured set does not reduce $\chi^2$. Each measured coefficient contributes more to $\chi^2$ than the largest non-measured coefficient. As in the low noise case, this is easily feasible if the true value of the coefficients is known.

\begin{figure}[tb]
  \centering
  \includegraphics{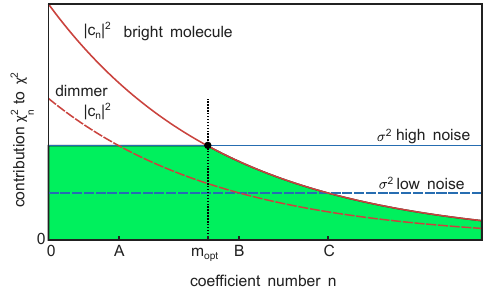}

  \caption{\textbf{Concept of adaptive sampling.} Each measured coefficient contributes to $\chi^2$ with a constant measurement uncertainty $\sigma^2$ (blue), while each unmeasured coefficient contributes with its true value $|c_n|^2$ (red). The total mean squared error is the area under both curves (green). The optimal choice  $m_\text{opt}$ occurs where the two curves cross. For other signal amplitudes, noise levels or integration times (as $\sigma^2 \propto 1/T_\text{int}$), the crossing point shifts (points A,B,C). Adaptive sampling should follow the optimal choice $m_\text{opt}$ regardless of the experimental conditions.}
   \label{sketch}
\end{figure}

Figure \ref{sketch} illustrates this concept. As we will see below, the absolute value of the coefficients $c_n$ for typical spectra can be approximated by an exponential decay with increasing $n$. The contribution to $\chi^2$ by the measured coefficients is given by $\sigma^2 m_\text{opt}/N_\text{tot}$, which is equal for all coefficients (blue curve). 
The contribution to $\chi^2$ by the non-measured coefficients is simply their absolute value squared (red curve), as shown here as a hypothetical example. The total mean squared error is the area under the curves (green). The optimal choice of $m_\text{opt}$ occurs where the two curves cross. For small coefficients, it is better not to measure them than to measure them with large uncertainty. 

As time progresses and the molecule has not yet bleached, the crossing point shifts (blue dashed curve). It also depends on the overall nonlinear signal strength and varies from molecule to molecule (red dashed curve). Therefore, we need an algorithm that can predict the absolute values of the coefficients based on those that have already been measured, and adapt the measurement scheme accordingly.

\section{Optimal sampling algorithm}

\subsection{The optimal window}

The adaptive algorithm must predict the absolute value of coefficients that have not yet been measured, or not with sufficient accuracy. Based on this prediction, it determines the crossing point $m_\text{opt}$ in Fig.~\ref{sketch} and which coefficients to measure next. The value of $m_\text{opt}$ will be uncertain and will fluctuate.  There will be situations in which $m_\text{opt}$ is reduced, meaning that it would have been better not to have measured certain coefficients in the first place. Handling these cases complicates the adaptive algorithm when it requires identical $N_n = N_\text{tot}/m_\text{opt}$ for all coefficients. Incorporating a window function that results in non-uniform sampling numbers $N_n$ makes the algorithm easier to implement.

A window function is a  real-valued factor $0 < f_n < 1$ that is multiplied on all Fourier coefficients $c'_n$ before the Fourier transformation. This is equivalent to convolving the spectrum $S'(\omega)$ with a kernel $F$ that is the Fourier transform of the $f_n$. A window function is always applied! If it is not mentioned explicitly, a rectangular window is used that is equal to one when $N_n > 0$ and zero otherwise. Would $\chi^2$ between the measured and true spectra decrease with a different window function? Is there a better function than a rectangle?

In statistics,  multiplying the measured $c'_n$ by $0 < f_n < 1$ is known as shrinkage\cite{Efron16}. This leads to
\begin{equation}
  \chi^2_n = \braket{\left| c_n - f_n c'_n \right|^2} = (1-f_n)^2 \, |c_n|^2 + f_n^2 \, \frac{\sigma^2}{N_n} \quad .
\end{equation}
which is the optimal Wiener filter\cite{Kailath00}.
We thus reduce the contribution of the measured coefficients ($\sigma^2$) a bit and pay by adding $|c_n|^2 $. It depends on their relative amplitude if this improves $\chi^2$. The equation has a minimum at 
\begin{equation}
   f_n = \frac{|c_n|^2 }{|c_n|^2 +  \sigma^2 / N_n} 
  \label{eq:optimal_window}
\end{equation}
 with a value of 
\begin{equation}
  \chi^2_n =   \frac{|c_n|^2 \sigma^2 }{N_n |c_n|^2 +  \sigma^2 } = \, 
  f_n \frac{\sigma^2 }{ N_n } < \frac{\sigma^2 }{ N_n } \quad .
\end{equation}
There is thus an optimal window function $f_n$ as defined by eq.~\ref{eq:optimal_window} that reduces $\chi^2$ between measurement and reality by the same factor. This value $f_n$ depends on the true value of the coefficient $c_n$ and the integration time $N_n$. When we allow a non-uniform window function, the optimal $N_n$ will thus differ between the coefficients.

\subsection{Ideal sampling scheme}

To investigate the ideal sampling scheme, we imagine a Daemon that knows the true values of the coefficients $c_n$. The Daemon would  proceed stepwise. After each measurement, the Daemon calculates the expected reduction $\Delta_n$ in $\chi^2$ when coefficient $n$ would be measured one more time
\begin{equation}
  \Delta_n = \frac{ |c_n|^2 \sigma^2} { |c_n|^2  N_n + \sigma^2 } - 
		\frac{ |c_n|^2 \sigma^2} { |c_n|^2  (N_n + 1) + \sigma^2 } \quad .
        \label{eq:Delta_n}
\end{equation}
The Daemon then selects the coefficient $n$ with the largest $\Delta_n$, measures at this delay and increments the corresponding $N_n$. This is repeated until the total number of measurements is reached, i.e. $\sum_{n} N_n = N_\text{tot}$.
After reaching the total number of measurements, the Daemon averages the measured coefficients at each delay and calculates the optimal window function according to eq.~\ref{eq:optimal_window} to get the optimal coefficients
\begin{equation}
  c_n' = f_n \bar{c}'_n \quad .
\end{equation}

It is instructive to look at the resulting window $f_n$ and integration times $N_n$. We find
\begin{equation}
  f_n = \frac{N_n}{A} = 1 - \frac{\sigma^2}{A |c_n|^2} 
\end{equation}
where the valid range of $n$ is limited to ensure $f_n \ge 0$ and $A$ is a suitable constant to guarantee $\sum_{n} N_n = N_\text{tot}$. Window $f_n$ and integration time $N_n$ are proportional. Assuming that the $|c_n|^2$ drop approximately by $\exp(-2 \gamma T)$ over one pixel, both $f_n$ and $N_n$  follow a step function with a smeared out edge, where the transition length is approximately $1/ \gamma T \approx 2$ pixels in our simulations.
Overall, the non-uniform optimised window function only reduces $\chi^2$ of the ideal sampling scheme by about 10\% (data not shown), but it will simplify the adaptive algorithm.

\subsection{The adaptive algorithm}

A Daemon knows the true values of the coefficients and can thus decide whether measuring a new one would be worthwhile. The experimenter does not have this knowledge. Since an absorption spectrum can be approximated by the sum of individual Lorentzians, the absolute value of the Fourier coefficients $|c_n|$ can be approximated by an exponential decay.
We thus propose to fit a simple single exponential decay as surrogate model $u^2$ to estimate the true values $|c_n|^2$ of the coefficients
\begin{equation}
        u^2_{n} =  u^2_{n_1, n_3} = A e^{ - 2 \gamma T (n_1 + n_3 + 1) } 
        \label{eq:surrogate_model}
\end{equation}
where the amplitude $A$ and decay constant $\gamma$ are the free fit parameters. See Appendix \ref{appendix:beating} for a discussion of the limitations of this surrogate model. 

We use maximum likelihood estimation to obtain these parameters. The measured $|\bar{c'}_n|^2$ follow  a noncentral chi-square distribution with two degrees of freedom (see eq. \ref{eq:pdf_noncentral_chisq} in  Appendix \ref{appendix:MSE_of_complex_values}). The  log-likelihood of the model parameters $A$ and $\gamma$ then becomes
\begin{equation}
   \mathcal{L}(A, \gamma) = 
  \sum_{n}
\left[
\log I_0 \left(\frac{2\sqrt{u^2_{n} |\bar{c'}_n|^2}}{\sigma^2 / N_n}\right)
- 
\frac{u^2_{n}}{\sigma^2 / N_n}
\right] \label{eq:log_a_gamma}
\end{equation}
where $I_0$ is the modified Bessel function of the first kind. Terms independent of $A$ and $\gamma$ have been omitted.

The adaptive sampling algorithm uses the surrogate model $u_n$ for the Daemon's true coefficients $c_n$.  Initially, the three coefficients with the lowest $t_1 + t_3$ are measured once to obtain starting parameters for the exponential fit. The following steps are then repeated iteratively until photobleaching occurs:
\begin{enumerate} \setlength{\itemsep}{0pt}
    \item Estimate $A$ and $\gamma$ by maximizing the objective function in Eq.~\eqref{eq:log_a_gamma} on a fixed grid of $A$ and $\gamma$ values.
    \item Using the estimated $A$ and $\gamma$, calculate the expected reduction $\Delta_n$ for each coefficient via Eq.~\eqref{eq:Delta_n}, employing the surrogate model $u^2_n(A,\gamma)$ as an estimate for the true coefficients $|c_n|^2$.
    \item Measure the coefficient corresponding to the largest expected reduction and increment $N_n$ accordingly.
\end{enumerate}
Upon photobleaching, all measured coefficients at each delay are averaged. The optimal filter window is then calculated according to Eq.~\eqref{eq:optimal_window}, using the final surrogate model $u^2_n(A,\gamma)$ to approximate the true coefficients $|c_n|^2$. This yields the optimal adaptively sampled coefficients
\begin{equation}
  c_n' = f_n \bar{c}'_n .
\end{equation}

Noise in the measured data results in errors when predicting unmeasured coefficients. As the integration time increases, however, the predictions will improve. If the value of a coefficient is overestimated, only the next measurement of that coefficient is affected because subsequent measurements are based on better estimates.

\section{Simulated data}\label{sec:SimData}

\begin{figure}[]
  \includegraphics{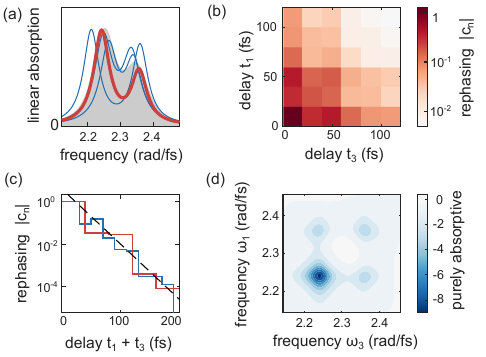}
   \caption{\textbf{Simulated molecules}. a) Linear absorption spectrum of the ensemble (gray area) and a few single molecules. The molecule used as an example in the following in plotted with a thicker line. b) Rephasing signal $|c_n|$ of the example molecule for short delays. c) Line cuts through b)   as function of $t_1 + t_3$ along $t_3  = 10$~fs (blue) and $t_1=t_3$ (red)  d) Purely absorptive spectrum (real part) of the example molecule.
}
   \label{fig_mol_spectra}
\end{figure}

In order to test the algorithm, we need the ground truth. Therefore, we simulate a fluorescence-detected 2D spectrum and add shot noise to it. This is then used as input for the algorithm, which determines which coefficient to measure next. We then compare the measured coefficients with the ground truth and calculate $\chi^2$. 

We use the analytical model by Tekavec et al. \cite{Tekavec2007} of the rubidium D lines, but adjust the parameters to resemble the V-level scheme of the light harvesting complex LH2 in purple bacteria \cite{vanOijen1999,Kunsel2019}. The parameters and their distributions are given in table \ref{tab:simulated_molecules}. Fig.~\ref{fig_mol_spectra}a shows the linear absorption spectrum of a few single molecules and the ensemble average. We vary the first and third inter-pulse delays ($t_1$ and $t_3$) and keep the population time $t_2 = 10$~fs fixed. We sample at $t_i = (n_i +1/2) T$ with $n_i = 0, 1, 2, \dots$ and  $T = 1 / 2\Omega = 20$~fs  so that all pixels are fully in the first quadrant. The rotating frame frequency is set to $\omega_{rot} = 2.3$~rad/fs. We scale the Fourier coefficients such that the absolute value of the lowest coefficient is one.

\begin{table}
  \begin{tabular}{l|ll}
parameter & mean & std. deviation \\
\hline
  $\omega_1$ (rad/fs) & 2.35  & 0.02 \\
  $\omega_2$ (rad/fs) & 2.25 & 0.02  \\
  $\mu_1$  & 1.0 & 0.2 \\
  $\mu_2$  & 1.3 & 0.2 \\
  $\gamma_0$ (rad/fs) & 0.03 & 0.005 \\
  $\tau_{fl}$ (ns) & 1 & 0 \\
  \end{tabular}
  \caption{\textbf{Parameters of the simulated molecules}. Transition frequencies $\omega_i$, dipole moments $\mu_i$ and dephasing rate $\gamma_0$ are  drawn from  normal distributions. The excited state lifetime  $\tau_{fl}$ is fixed.}
  \label{tab:simulated_molecules}
\end{table}

We calculate both the rephasing and the non-rephasing signal, but only use the rephasing signal for the adaptive algorithm. We find that the absolute value of the coefficients drops exponentially with $t_1+t_3$ (Fig.~\ref{fig_mol_spectra}c), supporting the choice of our surrogate model in Eq.~\eqref{eq:surrogate_model}. For a single molecule, the decay constant $\gamma$ corresponds to the line width of the narrowest spectral feature and is about 0.03~rad/fs in our simulations. Note that multiple spectral peaks lead to  a temporal beating, which can not be neglected when the ratio of peak separation $\delta$ to line width $\gamma$ becomes large (see Appendix \ref{appendix:beating}). The simulations use $\delta = 0.05 \text{\ rad/fs} \approx \gamma = 0.03 \text{\ rad/fs}$. Fig.~\ref{fig_mol_spectra}d shows the purely absorptive 2D spectrum (real part) of this single molecule.

\section{Results}

\subsection{\texorpdfstring{$\chi^2$}{Chi-squared} as function of integration time}

\begin{figure}
  \centering
  \includegraphics{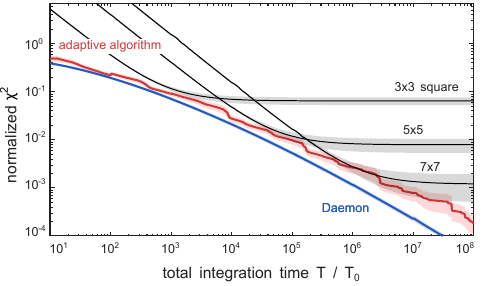}
  \caption{\textbf{$\chi^2$ as function of integration time.} The proposed adaptive algorithm (red) performs better than any typical square pattern (black) and almost as good as the ideal Daemon (blue). The shaded intervals indicate $\pm 0.25$ the standard deviation over the ensemble of molecules.  }
    \label{fig_chi3}
\end{figure}

\begin{figure*}
  \centering
  \includegraphics{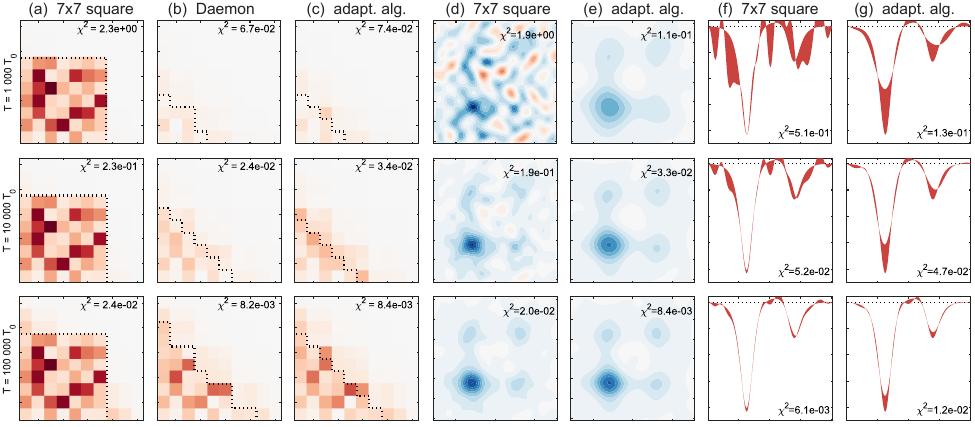}

    \caption{\textbf{Comparing the sampling schemes for one exemplary molecule}. Each row corresponds to a different total integration time $T / T_0 = 10^3, 10^4, 10^5$.
    \textit{Columns a-c: }
    Squared time-domain residuals ($\chi^2_n$)  using the   $7 \times 7$ square, the ideal  Daemon, and the adaptive algorithm. The resulting $\chi^2$ is given as inset, and varies strongly. The color scale is identical in each row and decreases from one row to the next by about a factor of $\sqrt{10}$. The dotted line encloses the measured coefficients.
        \textit{Columns d-e:} Corresponding real part of the purely absorptive 2D spectra. The Daemon is omitted as it visually agrees with the adaptive algorithm. 
            \textit{Columns f-g:} Corresponding line-cuts along the diagonal. The shaded area is the difference between the measured and the true curve.
}
    \label{fig_compare_both_at_N}
\end{figure*}

We perform a Monte Carlo simulation of 1000 single molecules. For each molecule, we calculate the time-domain 2D spectrum and apply different sampling schemes. We estimate the expected $\chi^2$ between the measured and true coefficients using Eq.~\ref{eq:MSE_target}, which depends on the sampling counts $N_n$. For the adaptive algorithm, we simulate noisy measurements, fit the surrogate model, and update the sampling counts after each measurement. We then evaluate $\chi^2$ analytically from the resulting $N_n$ values rather than from the simulated coefficient values. This reduces the computational effort required to obtain stable statistics.

We define a fundamental integration time $T_0$ such that the lowest coefficient has a signal-to-noise ratio $|c_0|/\sigma$ of one, where $\sigma$ is the RMS amplitude of the complex measurement noise. We simulate effective integration times up to $10^9 \, T_0$. The results are illustrated in Fig.~\ref{fig_chi3}. We normalize $\chi^2$ such that a value of one corresponds to measuring no coefficients at all, i.e., setting all coefficients to zero.

A common measurement scheme is to measure the lowest $K \times K$ time-domain points repeatedly until the molecule bleaches. In Jana et al. \cite{Jana2024} we used $K=7$. Two parts contribute to $\chi^2$: $\chi^2_n$ of the measured points given by $\sigma^2$, and $\chi^2_n$ due to the non-measured points given by the value of the coefficients as $|c_n|^2$. The first drops linearly with integration time, the latter remains constant (as soon as each point was measured once) and thus dominates for a large number of repeats. This is shown for $K=3, 5, 7$ in Fig.~\ref{fig_chi3}. 

The ideal case, the Daemon, is shown as blue curve. Its $\chi^2$ drops steadily with increasing integration time, as it always measures the coefficient that reduces $\chi^2$ most. The square scheme is always worse than the Daemon. It approaches the ideal case for a small range of effective integration times, but is also there about a factor of two worse.

Our adaptive algorithm (red) performs almost as good as the Daemon for small integration times, but begins to deviate as the integration time increases. This can be explained by the beating of the true time-domain signal, and is discussed in Appendix \ref{appendix:beating}. The periodically changing slope of the curve is also a result of the simple surrogate model. All coefficients at $t_1 + t_3 = const.$ become relevant for measurement simultaneously. The algorithm therefore oscillates between measuring only the old and only the new coefficients, which reduces $\chi^2$ in different ways. Nevertheless, even at large integration times, the adaptive algorithm does not perform worse than the best fixed square pattern. Only within narrow parameter intervals are the adaptive algorithm and the square scheme comparable. The adaptive algorithm is thus able to adapt to the individual molecule and its noise level, and performs well at any instance of photobleaching.

\subsection{Interpretation}

It is instructive to look at the mean squared error per coefficient $\chi^2_n$ for one exemplary molecule (Fig.~\ref{fig_compare_both_at_N}a-c). We show  three different integration times $T / T_0 = 10^3, 10^4, 10^5$ and  give as inset the resulting $\chi^2$. The measured coefficients are enclosed by a white line. Outside this line, all measurement schemes show the same values $\chi^2_n = |c_n|^2$ .  For illustrative purposes, the same random shot noise has been used in all panels.

In the square scheme, measurement noise dominates for small values of $T / T_0$ . The residual exponential decay of the not measured coefficients would dominate for even larger values of $T / T_0$ (not shown). The value $\chi^2_n$ in the top right corner of the measured square (maximum $t_1+t_3$) is larger than its neighbours. Therefore, it would have been better not to take a measurement here. The ideal Daemon and the algorithm measure in a triangular pattern, following lines of $t_1+t_3 = \text{const.}$. The inside and outside of the triangle blend together, reducing the noise of the measured coefficients just enough to match that of the unmeasured coefficients. The difference of the Daemon to the adaptive algorithm arises from deviations from the pure triangular shape. The Daemon deviates slightly from this shape, whereas the algorithm is forced to maintain it, resulting in slightly lower performance.

\subsection{Comparison of spectra}

Until now, our discussion has focused on the Fourier coefficients of the rephasing signal. In the simulations, we base our decisions on the measured rephasing signal coefficients, while also measuring the non-rephasing signal at the same delay. Combining the Fourier transforms of both signals yields the purely absorptive signal (see columns d--e of Fig. \ref{fig_compare_both_at_N}). A cut along its diagonal corresponds to the linear absorption spectrum. We have omitted the ideal Daemon, as there is no visible difference compared to the adaptive algorithm. The noise-free spectrum has already been shown in Fig. \ref{fig_mol_spectra}d.

The spectra shown provide a visual indication of the noise present.  The $\chi^2$ values (shown in the inset) of the 2D spectra are consistent with those of the time-domain rephasing signal. For line cuts along the diagonal, the $\chi^2$ of the adaptive algorithm also aligns with that of the time domain. However, the square scheme performs better than expected based on the time-domain $\chi^2$. This is because it measures the high-frequency components from the outset, helping to capture the width of the peak in the spectrum.  This comes at the expense of increased  noise in the background.

It becomes obvious that the choice of the optimization metric has an influence on the result. We chose to aim for a faithful reconstruction of the full spectrum. If one were only interested in the peak position or width, a different metric would lead to a different optimal sampling scheme.

\section{Conclusion}

When the signal-to-noise ratio is low, it is better to measure fewer coefficients over a longer integration time than to measure more coefficients over a shorter integration time. Ideally, by the end of the measurement, the absolute value of the smallest coefficient should just exceed the noise level. In a fixed square pattern, this can only be achieved over a small range of parameters. In many cases, a triangular pattern is preferable, as its shape reflects the exponential decay of the coefficients.

Selecting the optimal coefficients beforehand becomes impossible if the total integration time or strength of the nonlinear signal cannot be predicted due to factors such as photobleaching or variation between molecules. We have developed an adaptive algorithm that can handle these cases. It estimates the absolute value of the Fourier coefficients based on previous measurements and then determines which coefficients to measure next. The algorithm can adapt to individual molecules and their signal-to-noise ratio and performs well at any instance of photobleaching. Provided the noise amplitude scales with the square root of the integration time, the algorithm can be applied, e.g., to increase the sample throughput in an ensemble experiment.  In our previous experiment (Jana et al. \cite{Jana2024}) the effective integration time in units of $T_0$ was between 100 and 400. As can be seen by comparing with Fig.~\ref{fig_chi3}, the $7 \times 7$ square pattern was far from optimal.  We expect to reduce $\chi^2$ by approximately a factor of 30 using the adaptive algorithm. Part of this improvement could be invested to measure at different population times $t_2$.

We have demonstrated the performance of the adaptive algorithm using simulated data, and we are confident that it will perform well in real-time analyses to make \textit{ad hoc} decisions about the next measurement point. The first experiments are currently underway. We also believe that it will be useful for other high-noise Fourier domain experiments based on electron emission \cite{Uhl:21}, electric current \cite{Xiang2024}, or optical near-fields \cite{Esses2026}. Our  goal was to acquire the full 2D spectrum. Reducing the measurement to only a few descriptive values, such as peak width or position \cite{Bolzonello2024,Wang2020}, or using pulse shapes other than identical Gaussians \cite{Roeding2017}, could further reduce the integration time.

\appendix

\section{Time-domain beating of the signal} \label{appendix:beating}

In the adaptive algorithm, we employ a single-sided exponential decay as a surrogate model to predict time-domain coefficient amplitudes from previously measured values. This choice is motivated by the Fourier transform of a Lorentzian lineshape, which corresponds directly to a single-sided exponential decay. However, the presence of two or more peaks in the spectrum introduces an interference beating pattern in the time-domain signal. Assuming two identical Lorentzians centred symmetrically at frequency offsets $\pm\delta$ relative to the rotating frame frequency, the time-domain signal is given by
\begin{equation}
  s(t) = 2A \, e^{-\gamma t} \, \cos(\delta t) \quad , 
\end{equation}
where $\gamma$ represents the decay rate. The influence of this beating on the signal envelope is determined by the ratio of the peak decay rate to the frequency offset, $\gamma / \delta$. When the peak separation is large compared to the decay rate ($\gamma \ll \delta$), the oscillation frequency is high relative to the decay timescale $\tau = 1/\gamma$. Consequently, beating nodes manifest at short delay times $t$. Conversely, as the peak separation decreases ($\gamma \gg \delta$), the beating period exceeds the signal lifetime, and the time-domain signal approaches a single exponential decay.

The impact of this beating on the sampling strategy becomes evident when comparing the coefficients sampled by the adaptive algorithm with those selected by an ideal Daemon. Because the Daemon possesses full prior knowledge of the true coefficients, it accounts for the beating pattern and systematically bypasses sampling points near signal nodes where the amplitude is suppressed. In contrast, the adaptive algorithm relies on a monotonic exponential surrogate model. Being oblivious to the beating nodes, it redundantly samples these low-amplitude coefficients. This effect can be mitigated by narrowing the permitted range of values for the decay constant $\gamma$ when fitting the surrogate model. Fixing $\gamma$ at the ensemble mean value (Tab.~\ref{tab:simulated_molecules}) results in a $\chi^2$ that closely resembles the ideal Daemon shown in Fig.~\ref{fig_chi3}.

This discrepancy between optimal sampling and the adaptive algorithm is illustrated in Fig.~\ref{fig_compare_both_at_N}, where the Daemon  excludes coefficients, whereas the adaptive algorithm continues to sample them. Consequently, as shown in Fig.~\ref{fig_chi3}, the performance of the adaptive algorithm deviates from the ideal Daemon at longer total integration times. As longer integration times extend measurements to larger delays $t$, neglecting the beating characteristics in the surrogate model leads to an increasing number of redundant measurements on small-amplitude coefficients that were skipped by the ideal Daemon. Therefore, systems with pronounced peak separation ($\gamma \ll \delta$) and long total integration times necessitate a surrogate model that explicitly incorporates multi-peak interference.

\section{Note on the variance of complex values} \label{appendix:MSE_of_complex_values}

The Fourier coefficients $c_n = a + i b$ and their measured values $c'_n = a' + i b'$ are complex-valued. 
We define $\sigma^2$ as the mean squared modulus of the complex measurement error, $\left\langle |c_n - c'_n|^2 \right\rangle = \sigma^2$. 

As we derived in the supplementary material of Jana et al. \cite{Jana2024}, the variance of each measured component $a'$ and $b'$ (assuming fixed true values $a$ and $b$) is related by shot noise to the total number of photons $N_{photon}$ that contribute to the measurement of this coefficient, i.e. $\sigma^2/2 = 2 / N_{photon}$.
\begin{equation}
a'-a,\ b'-b \overset{\text{indep.}}{\sim}
\mathcal N\left(0,\frac{\sigma^2}{2}\right).
\end{equation}
We thus simulate measurements by adding independent normally distributed random numbers of mean zero and variance $\sigma^2/2$ to each of the true complex components $a$ and $b$.

The term $|c_n - c'_n|^2$ has a mean of $\sigma^2$ and a variance of $\sigma^4$.

The absolute value squared $|c'_n|^2$ follows  a scaled noncentral chi-square distribution with two degrees of freedom with probability density\cite{Proakis08}
\begin{align} \label{eq:pdf_noncentral_chisq}
&p \Bigl( |c'_n|^2 \, \Bigl| \, |c_n|^2, \sigma^2  \Bigr) = \\
&\frac{1}{\sigma^2 }
\,
\exp\left[
-\frac{|c'_n|^2 + |c_n|^2}{\sigma^2}
\right]
\,
I_0
\, 
\left(
\frac{2\sqrt{|c'_n|^2 |c_n|^2}}{\sigma^2}
\right). \nonumber
\end{align}
where $I_0$ is the modified Bessel function of the first kind. It has a mean of $|c_n|^2+\sigma^2$.

The absolute value  $|c'_n|$ follows a Rice distribution with a probability density\cite{Proakis08}
\begin{align}\label{eq:pdf_Rice}
&p\Bigl( |c'_n|  \, \Bigl| \, |c_n|,\sigma^2 \Bigr) = \\
& \frac{2 |c'_n| }{\sigma^2}
\exp\left[-\frac{|c'_n|^2+|c_n|^2}{\sigma^2}\right]
I_0\left(\frac{2\, |c'_n| \, |c_n|}{\sigma^2}\right). \nonumber
\end{align}
with a mean of 
\begin{align} \label{eq:mean_Rice}
&\left\langle |c'_n| \right\rangle  = \\
& \sigma\frac{\sqrt{\pi}}{2}
e^{-\rho^2/2}
\left[
(1+\rho^2)I_0\left(\frac{\rho^2}{2}\right)
+
\rho^2 I_1\left(\frac{\rho^2}{2}\right)
\right], \nonumber
\end{align}
where
$ \rho= |c_n| / \sigma$ and  where $I_0$ and $I_1$ are modified Bessel functions of the first kind.

\section*{Acknowledgements}

We acknowledge the financial support of the German Science Foundation (DFG) via IRTG OPTEXC and project 524294906.

 \section*{Author declarations}

 \subsection*{Conflict of Interest} The authors have no conflicts to disclose.

 \subsection*{Author Contributions}
\textbf{Sebastian Stadler:} Investigation (equal), Software (support), Visualization (equal), Writing (equal).
\textbf{Simon Kehrer:} Investigation (support),  Visualization (support), Writing (support).
\textbf{Thorsten Schumacher:} Investigation (support), Visualization (support), Writing (support).
\textbf{Markus Lippitz:} Conceptualization, Investigation (equal), Software (lead), Writing (lead).

\section*{Data Availability} 

The data that support the findings of this study are openly available on GitHub at \url{https://github.com/Lippitz-Lab/adaptive_sampling_simulations}.


\section*{References}
\bibliography{references}

\end{document}